\documentclass[pdflatex,sn-mathphys]{sn-jnl}% Math and Physical Sciences Reference Style
\jyear{2023}%

\theoremstyle{thmstyleone}%
\theoremstyle{thmstyletwo}%

\theoremstyle{thmstylethree}%

\begin{document}

\title[Article Title]{The aesthetics of cyber security: How do users perceive them?}


\author*{\fnm{Mark} \sur{Quinlan}}\email{mark.quinlan@cs.ox.ac.uk}

\author{\fnm{Aaron} \sur{Ceross}}\email{aaron.ceross@cs.ox.ac.uk}

\author{\fnm{Andrew} \sur{Simpson}}\email{andrew.simpson@cs.ox.ac.uk}

\affil{\orgdiv{Department of Computer Science}, \orgname{University of Oxford}, \orgaddress{\street{Wolfson Building, Parks Rd}, \city{Oxford}, \postcode{OX1 3QD}, \state{Oxfordshire}, \country{United Kingdom}}}


\abstract{While specific aesthetic philosophies may differ across cultures, all human societies have used aesthetics to support communication and learning.  Within the fields of usability and usable security, aesthetics have been deployed for such diverse purposes as enhancing students' e-learning experiences and optimising user interface design.  In this paper, we seek to understand how individual users perceive the visual assets that accompany cyber security information, and how these visual assets and user perceptions underwrite a distinct \emph{cyber security aesthetic}.  We ask, (1) \emph{What constitutes cyber security aesthetics, from the perspective of an individual user?} and (2) \emph{How might these aesthetics affect users' perceived self-efficacy as they informally learn cyber security precepts?} To begin answering these questions, we compile an image-set from cyber security web articles and analyse the distinct visual properties and sentiments of these images.}




\maketitle

\section{Introduction}

Visual media, like illustrations~\cite{lin2018impact} and diagrams~\cite{hattwig2013visual}, have accompanied text since the very first written documents~\cite{nichols1858illustrations}, providing clarification, communicating distinct emotions or opinions, and serving more subversive ends like propaganda~\cite{marland2012political,cooper2008war,meyer2008aesthetics}.
Today, digital technologies have introduced new types and ways of accessing visual media while rapidly integrating media consumption into the daily lives of large segments of the global population~\cite{david2010impact}.  Still, contemporary online news articles and blog posts are often accompanied by visual media that serve many of the same communicative purposes as those in the earliest human documents~\cite{mitchell2005just}.

%% introducing the concept of cybersecurity and why

The fields of usable security and cyber security have used visual media to support learning.  For example, user interface designers used (and later abandoned) skeuomorphism to generate easily identifiable visual objects that could help users to navigate new interfaces~\cite{page2014skeuomorphism,curtis2015rhetoric}, and they frequently use colours to draw attention to salient information and features within e-learning platforms~\cite{tharangie2008kansei,reyna2013importance}.  Meanwhile, usable security experts explore how visual cues can aid users outside of formal learning environments~---~that is, how and when visual media can facilitate \textit{informal learning}.  Informal learning is the primary way in which adults learn about the world around them~\cite{malcolm2003interrelationships, ollis2011learning}, and it typically occurs when individuals actively choose to seek out new ideas and advice.

%Informal learning where individuals learn something without that activity surrounding it is called \textit{incidental learning}~\cite{malcolm2003interrelationships, bull2008connecting}.  The primary difference in this case is intentionality: informal learning requires some kind of prior impetus; incidental learning does not and is often found as a by-product of carrying out other tasks~\cite{malcolm2003interrelationships}.

Understanding how visual media work to support communication and learning is a complex task, and it is divided amongst many scholarly disciplines.  First and foremost amongst these is the ancient philosophical branch of aesthetics, which concerns itself with the nature of perception, taste, and the values of sensory qualities (e.g., beauty).  In the context of this paper, aesthetics entail the perceptual logic that allows individuals to instinctively analyse meaning and appraise quality / truth when consuming visual media~---~whether as stand-alone objects, as part of a user interface, or as an accompaniment to text.  By extension, learned aesthetic preferences may influence how individuals navigate information or environments~\cite{flavian2009heuristic,joshi2011aesthetics,shires2020cyber,carroll2021usable}, and they may play a significant role in informal learning.

%--- where the aesthetics allow for instinctive analysis of meaning and truth that users undertake when consuming visual media, whether as stand-alone objects, as part of a user interface or as an accompaniment to text, thus aiding their capacity to learn informally~\cite{dobbs1998learning}.

There is a body of literature within the computer science usable security field that looks at users' aesthetic perceptions of technology.  For example, work by Fogg et al.~\cite{fogg2009behavior} found that almost half of all users used aesthetic judgements to infer the credibility of a site's content.  Compounding these results, Robinson et al.~\cite{robinson2020digital} found that individuals use aesthetic information to make rapid judgements about content, and Alsudani and Casey~\cite{alsudani2009effect} reported that these judgements occur within about 3.5 seconds.  In the sub-area of cyber security aesthetics, \cite{ma2006cyber} outlined various cyber security visualisation techniques, and \cite{shires2020cyber} looked at the adaptation of neo-noir aesthetics in cyber security visual media.  There is also work within the field of digital exhibits~\cite{bernal20201} and the transfer of security aesthetics into cyber security~\cite{ghertner2020futureproof}.  
%, which allows us to infer the use of cyber security aesthetics potential as a pedagogical tool of which the efficacy may not yet have been assessed.
Taken together, these precedents suggest that cyber security aesthetics can serve as a pedagogical tool, helping users to parse information and act upon it.  Thus, improving our understanding of these aesthetics could help us to improve the efficacy of security advice dissemination.
%makes choices dependent on it.

In this paper, we explore the following research aims:
\begin{enumerate}
\item \emph{What cyber security aesthetics consist of, from the perspective of an individual user.}
\item \emph{Provide an explorative discussion as to the manner in which these aesthetics may affect users' perceived self-efficacy as they informally learn cyber security precepts.}
\end{enumerate}

To do so, we report on how we assembled an image-set of cyber security images that reflects what a user typically sees within an informal learning environment.  The corpus spans 1,027 images and is derived from English language news and online magazine articles from the United States, Canada, and the United Kingdom.  The images are organised into several classes, which we derived by extracting visual information from the raw images and mapping them to semantically meaningful keywords, and performed a colour similarity analysis.  %We confirmed the internal consistency of each image class through colour analysis.

The remainder of the paper is organised as follows.  In Section~\ref{Background} we provide the background to, and the motivation for, the work described in this paper.  In addition, we define some of the terms of interest.  In Section~\ref{Methodology} we describe the process used to create the image-set, as well as the data cleaning process we used to develop the image-set into usable images.  In Sections~\ref{Results} and~\ref{Discussion} we present and discuss our results, placing them in a broader context.  Section~\ref{future} presents potential research directions for the broader research community.  Finally, Section~\ref{Conclusion} concludes the paper.

\section{Background and Motivation} \label{Background}

In this section we discuss the background to, and the motivation for, the work described in this paper.  As aesthetics is such a broad and sometimes ambiguous term~\cite{dewey1934art}, we begin by providing an overview of aesthetics research in Sections~\ref{Secsemiotics} and~\ref{Greeks}, establishing its relevance for our research aims.  We then consider the potential efficacy of aesthetics for cyber security in Section~\ref{efficacy}.  Section~\ref{aesthete} then returns to our overarching research aims.

\subsection{Aesthetics and meaning} \label{Secsemiotics}

In Ancient Greece, aesthetics were first described as a `sensation', or the ability to interact with external stimuli through our bodily senses~\cite{beardsley1975aesthetics}.  
%We can draw from the philosophical world, where
Later, Kant~\cite{Kant1892-KANTCO-3} espoused the importance of aesthetics for all human domains, arguing that, without its sense-making power, data would simply remain chaotic, lacking meaning and structure.\footnote{This could mean that data acquires a certain aesthetic once created (mathematical aesthetics~\cite{Kant1892-KANTCO-3}), or that, to have an aesthetic perspective, one must assemble data fragments into something meaningful regardless of outcome~\cite{Kant1892-KANTCO-3}.} However, if aesthetics help to render a \textit{shared} sensible reality, as asserted by Kant~\cite{Kant1892-KANTCO-3, zander2016intuition}, then aesthetic perception must be universal, narrative, and standardised.  This brings us into the sphere of semiotics~---~the study of signs, symbols, and symbolisation, or of the devices and practices that help to stabilise meanings.

For the purposes of this paper, we define aesthetics as the perceptual logic that allows individuals to instinctively analyse meaning in visual media, and semiotics as the conventionalised meanings arising from this perceptual process.  In other words, where aesthetic objects exist solely for their own purposes~\cite{Kant1892-KANTCO-3}, informing perceptible meaning~\cite{walsh1974aesthetic,wissenburg2012aesthetic} without \textit{requiring} a specific meaning to be understood, semiotic objects contain explicitly built-in meanings, whether skeuomorphic or otherwise~\cite{carroll2021usable,barbosa2021semiotics}, and can enhance the meaning of words they are associated with (for example, as a compendium to a body of text~\cite{dewey1934art}).

Both aesthetic and semiotic perspectives remain relevant in contemporary philosophical discourse, as well as in the practice of cyber security.  For example, \cite{doi:10.1080/10350330.2019.1587843} describe how viewers attempt to derive meaning from \textit{key referent objects} contained within an image.  Insofar as these objects are universally understood, they may yield what Ranciere calls `a shared sense of perception'~\cite{sayers2005jacques}.
Furthermore, in the context of cyber aesthetics, most interactions at the interface level are directed by symbols and imagery such as icons, pointers, image thumbnails~---~which themselves can contain semiotic objects~---~and so forth.  All of these objects help to convey a system logic to end users, thus enhancing accuracy and intuitiveness~\cite{rudner1951semiotic}.

Of course, in a nascent field like cyber security, aesthetic systems have not necessarily been formalised into stable semiotic resources.  As such, we must not preemptively constrain our analysis to specific image contents, addressing instead the full spectrum of aesthetic objects relevant to cyber security communication.  In this case, we define these to be \textit{(visible) digital image-objects that may themselves contain semiotically legible signs, and which have been added as an adornment or supplement to relevant cyber security literature}.  Although this definition presents some limitations (discussed in due course), it allows us to account for the narrative functions of aesthetics, as well as for its use in learning.

%What remains to be seen is how these assumptions, made about aesthetics in a general sense, translate to what we find in the current paradigm for cyber security aesthetics.  The following Sections aim to help us in this search.

%Bridging the gap between those early definitions and today is a plethora of words spanning the arts and sciences, with the nascent field of \textit{neuroaesthetics} (definition) one of most recent areas of research~\cite{pearce2016neuroaesthetics}.

\subsection{Aesthetics and learning}\label{Greeks}

Because we are primarily interested in how users may interpret cyber security aesthetics in an informal learning context, our understanding of aesthetics is informed by contributions such as that of \cite{chatterjee2014aesthetic}, which draws from the humanities and sciences to illustrate how aesthetics influence the choices humans make in their given domains of activity.  Earlier work by \cite{carper1975fundamental} went further, identifying aesthetics as one of four distinct structures human beings use when developing knowledge, the other three being personal, empirical, and ethical.  For Carper, the `knowing' of aesthetics takes the other three structures and enhances them into a new understanding, creating meaning from otherwise abstract works~\cite{carper1975fundamental}.  Building on Ancient Greek notions of aesthetics, \cite{keenan2016use} proposed the concept of `aesthetic knowledge', wherein sensory experiences form embedded relationships with phenomena such as colour and shape.  According to Keenan, when prior aesthetic knowledge is combined with information (or, in our case, images) from user interfaces and other elements within a digital experience, users can associate prior meanings with this new information and thereby generate unanticipated interactions~\cite{keenan2016use}.
 
Taken as a form of knowledge, aesthetic design can enhance users' ability to make effective decisions based on a mixture of intuition and explicitly learnt knowledge~\cite{carroll2010designing}.
We know that people often make decisions based on intuition rather than analytical inference, `sensing' a correct choice without being able to offer a logical explanation for it~\cite{zander2016intuition}; we may also expect that aesthetic objects can serve to stimulate this intuition.  For example, within human--computer interaction, supplementary visual assets that convey a feeling of uncertainty or ambiguity can help individuals to comprehend uncertainty even when it is not explicitly communicated in words~\cite{fernandes2018uncertainty}.  It follows that aesthetic knowledge will impact knowledge acquisition in any given field, including fields where many users rely on informally learnt knowledge (such as cyber security~\cite{rader2012stories}) or cases where decision makers do not have prior experience with the given situation~\cite{zander2016intuition}.

\subsection{Aesthetics and self-efficacy} \label{efficacy}

Clearly, the way in which we are presented with information visually impacts our understanding of, and subsequent decision making towards, a particular topic.  As such, usable security research, user interface design, and cognitive psychology theory have sought to better understand how and why users make aesthetic decisions, and how aesthetic attributes can be designed to achieve certain ends.  For instance, some scholars studying the ethics of technological development have proposed tools to help designers build fairness and transparency into digital libraries and interface designs through deliberate aesthetic planning~\cite{barbosa2021semiotics}.  Other researchers have explored how particular aesthetic / semiotic interpretations of user interfaces can enable users to complete a given task more efficiently~\cite{carroll2021usable}.  This latter effect is particularly interesting in the context of cyber security, given the brunt of responsibility that individual users have to bear for protecting themselves, their devices, and their networks online.

%As mentioned in Chapter \ref{intro}, many studies have concerned themselves the the efficacy potential of advice.  In many of these studies, they relate to the individual users' ability to practice \textit{self-efficacy}.

One important concept implicated in users' decision making is self-efficacy~---~a generative capability to organise one's skill-sets and beliefs towards a desired outcome~\cite{bandura1999self}.
According to self-efficacy theory, individual users implicitly judge their own ability to cope with a given situation, thus developing self-efficacy beliefs for a specific domain.  These beliefs inform whether individual users will initiate certain behaviours and carry them through to successful outcomes~\cite{maddux1995self, bandura1999self}.  Furthermore, self-efficacy is closely related to motivation: the greater the challenge a user faces, the more self-efficacy they will need to sustain their motivation~\cite{bandura1999self, stumpf1987self}.  Because cyber security is perceived to be both important and complex, users tend to exhibit limited self-efficacy in this domain (as explained by \cite{Herley} and explored by \cite{halevi2016cultural} through psychological and cultural means).

Many people develop some degree of self-efficacy through their identification with role models: people similar to themselves who display, and thereby make accessible, certain aspirational attitudes, behaviours, or capacities~\cite{ajzen1991theory}.  For example, \cite{bosma2012entrepreneurship} observed that role models in the media can encourage entrepreneurship amongst their viewers.  Applying these insights to cyber security aesthetics, we may suggest that researchers can utilise aesthetics to enhance users' self-efficacy by providing models, structuring and directing behaviour towards goal setting, and measuring progress towards these goals~\cite{carroll2010designing}.

\subsection{Our expectations for this exploration} \label{aesthete}

%[frame expectations for research question answers, given what has been presented in this Background section]

To summarise, we expect that users acquire an aesthetic literacy when they are repeatedly exposed to domain-specific content, and that this literacy helps them to navigate and derive meaning from future content.  As per our first research aim, we aspire to understand the aesthetics (and thus aesthetic literacies) operative in the domain of cyber security, and so we will imaginatively replicate the process whereby users develop these literacies~---~that is, repeated exposure to the aesthetic objects of cyber security~---~by compiling an image-set of cyber security's primary aesthetic objects, allowing us to appraise and compare them at once.  We have defined these objects to be images that may themselves contain legible signs, and which are typically part of a larger piece of content like an online article.  As per our second research aim, we will interpret the resulting aesthetics in terms of their likely effects for users' self-efficacy.

%[list requirements needed to test / verify hypotheses / expectations and answer research questions here~---~also, explain why annotating an image-set of just cyber security-related images is a sufficient way to answer these questions (as opposed to comparing cyber security images with other, but perhaps similar, disciplinary artifacts to identify differences; interviewing users, cyber security experts, illustrators, article writers, etc. about their interpretative and communicative / meaning-making processes and expectations; performing A / B testing re: image efficacy; etc.)]

%https://perma.cc/Y4NR-ZK2C

%WJT Mitchell famously wrote that “all media are mixed media” (2005: 260).

%DUDE, Shires2020 has a great amount of info on cyber asethetics.

%These more contemporary notions of the aesthetic 

\section{Methodology} \label{Methodology}

In this section we discuss the research design of our study, which proceeded in five steps:
\begin{enumerate}
    \item developing the image-scraping tool in Python to extract images from structured data sources; 
    \item configuring a viable search methodology based on common cyber security terminology; 
    \item cleaning the initial pool of images to yield a usable image-set; 
    \item preparing the labels and resources needed for computational image classification; and 
    \item performing colour analysis to confirm the internal consistency of each image class.
\end{enumerate}

\subsection{Developing the image-scraper}\label{scraper}
 
Web scraping is a popular digital research technique that allows researchers to automatically capture freely available online data~---~that is, data that does not require privileged access~\cite{marres2013scraping}~---~via the use of scrapers.  Our image-scraper is a simple tool designed to capture images from pages selected by our search methodology (discussed in Section~\ref{search}).  Rather than incorporate additional system logic to ensure that all images were viable candidates for analysis, we chose to refine the image-set through subsequent data cleaning (discussed in Section~\ref{cleaning}).

\subsection{Deriving relevant images from search terms} \label{search}

To establish the list of search terms needed to guide the image-scraper, we followed the precedent of \cite{schatz2017towards}, who used Google Trends to automatically collect real search terms employed by the target audience.\footnote{\cite{schatz2017towards} sought to derive a more precise definition of security, and so they collected the terms that individuals used to search for security content.}  This focus on user-centred definitions excluded the possibility of replicating the work of \cite{humayun2020cyber}, who looked at primary studies undertaken within academia.  Instead, we followed the \textit{Systematic Mapping Study} protocol presented by \cite{kosar2016protocol}.

We defined a set of base search terms (for example, `cybersecurity' \texttt{OR} `cyber' \texttt{AND} `security') and then added search terms derived from Google Trends (online OR advice OR protection OR protect OR prevent OR preventative OR tips OR email OR social network OR password OR hack OR hacked OR hacking).  All search terms were technology-agnostic~---~they did not include explicit references to specific products or services.  The image-scraper then returned all images that corresponded with content that included these terms within the title or body text.  Though not exhaustive, this strategy yields an image-set that adequately represents operative definitions of cyber security, as actualised by users.  There is, of course, scope for future improvement.

\subsection{Cleaning the data} \label{cleaning}

The aforementioned search strategy yielded an initial image-set of 4,784 images, which we then subjected to an initial data cleaning based on the following inclusion / exclusion criteria (to enable consistency):

\begin{itemize}
    \item The image must be derived from a news or blog article that directly addresses at least one aspect of cyber security and / or explicitly contains our search terminology.  Blog articles were limited to tutorials, editorials, tool demonstrations, and discussions of technical reports.  Due to the nature of the assessment and the search methodology, we only retrieved images from English-language sources.
    \item The image must be accessible and not hidden behind a paywall or other kind of lockout mechanism, as these obstacles restrict the amount of text that can be retrieved, making it difficult to explain why some images were included in a given article or blog post (that is, the role that the images serve in relation to the text).
    \item The image cannot be a corporate logo or advertisement (like the lead slide of a corporate presentation).
    \item The image must be at least 360x640 pixels for ease of processing.
    \item The image must be in either .jpg or .png format.
\end{itemize}

Applying these criteria, we reduced the initial pool of 4,784 images to 3,757 usable images.  We then counted and removed all duplicates\footnote{We counted the number of duplicates to assess the extent of duplication within cyber security aesthetics.} and then down-sampled our images to a standard pixel resolution.  This yielded a final image-set of 1,027 individual images, which we used for analysis \footnote{This image-set can be found here: \url{https://huggingface.co/datasets/Quinm101/cyberaesthetics}.}.

\begin{figure}
    \centering
    \includegraphics[scale=0.2]{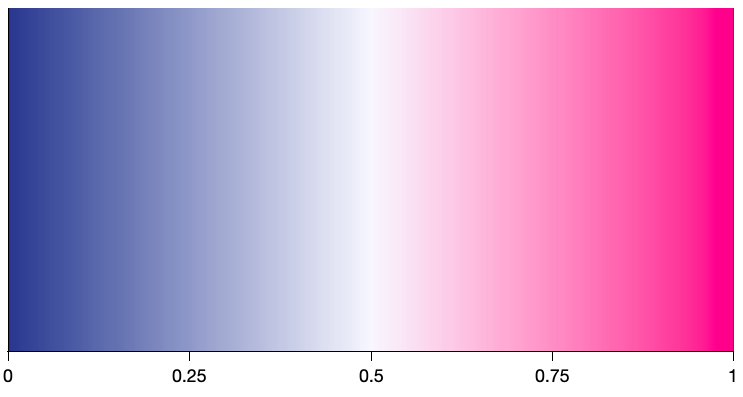}
    \caption{The colour distance charts for our image class heat-maps.  Similarity decreases as the x-axis moves from 0 (blue) to 1 (pink).}
    \label{fig:colourdistance}
\end{figure}

\subsection{Classifying the images} \label{sec:data}

The next step in the process involved feature extraction~---~a form of quantitative image classification wherein categorising labels are assigned to images based on specific extracted features.  For our image-set, we chose to begin with object recognition to identify any potential semiotic objects (or signs) before moving onto semantic categorisation (categorising emotion or other subjective features).

% We initially attempted to instigate unsupervised learning through $k$-means (UFLK), following the example of \cite{gultepe2018predicting}, who utilised this method in conjunction with two others to ascribe certain styles of painting to classical fine art.  However, the varied nature of the image-set made automated feature extraction untenable, resulting in unsatisfactory classification performance (as determined by manual verification of accuracy and F-score).

We utilised a variation of the Bag-of-Words (BoW) model to provide human-assigned classifications for our image-set.  The BoW model is often used in situations where images require text categorisation but word order is not particularly important.  We based our model on work by \cite{csurka2004visual}, selecting three knowledgeable cyber security researchers to manually locate dominant interest points in individual images (recreating feature extraction) and derive labels that represent these interest points.  This solution presented several advantages over an automated set-up, as the expert knowledge allowed for more concise labelling, the ability to label occluded objects that would have been missed by automated methods, and the ability to construct a clearly defined codebook for our classification labels based on prior expertise.  However, the contextual awareness these experts brought to the labelling exercise may have introduced some biases.  This limitation could be mitigated in future studies by recruiting a wider range of annotators.

\subsection{Measuring image similarity through colour} \label{colour}

To confirm the internal consistency of each image class derived from our classification process, we utilised Weller and Weastneat's~\cite{weller2019quantitative} quantitative, colour-based method for measuring image similarity.  This involved transforming each image's pixels into 3D coordinates to produce a multidimensional color histogram for each image, then using the earth mover's distance measure~\cite{rubner2000earth} to compute the pairwise distances between histograms. 
We opted for this method over contour-recognition for object classification (as used by \cite{gupta2019nose, waldchen2018plant}) because we had already classified our images according to their dominant features.  Colour similarity measures also allow us to more confidently make qualitative assessments relevant to our research aims.

Colour similarity heat-maps for each class will be shown later on in this paper, and can be interpreted through Figure \ref{fig:colourdistance}.  Each heat-map represents the relationship any given image has to the other images within its class, with blue cell colours indicating greater similarity and red cell colours indicating lesser similarity.

%For the semantic categorisation

%In the next section, we can see the results and their associated discussion.

\section{Results} \label{Results}

\begin{figure}
    \centering
    \includegraphics[scale=0.118]{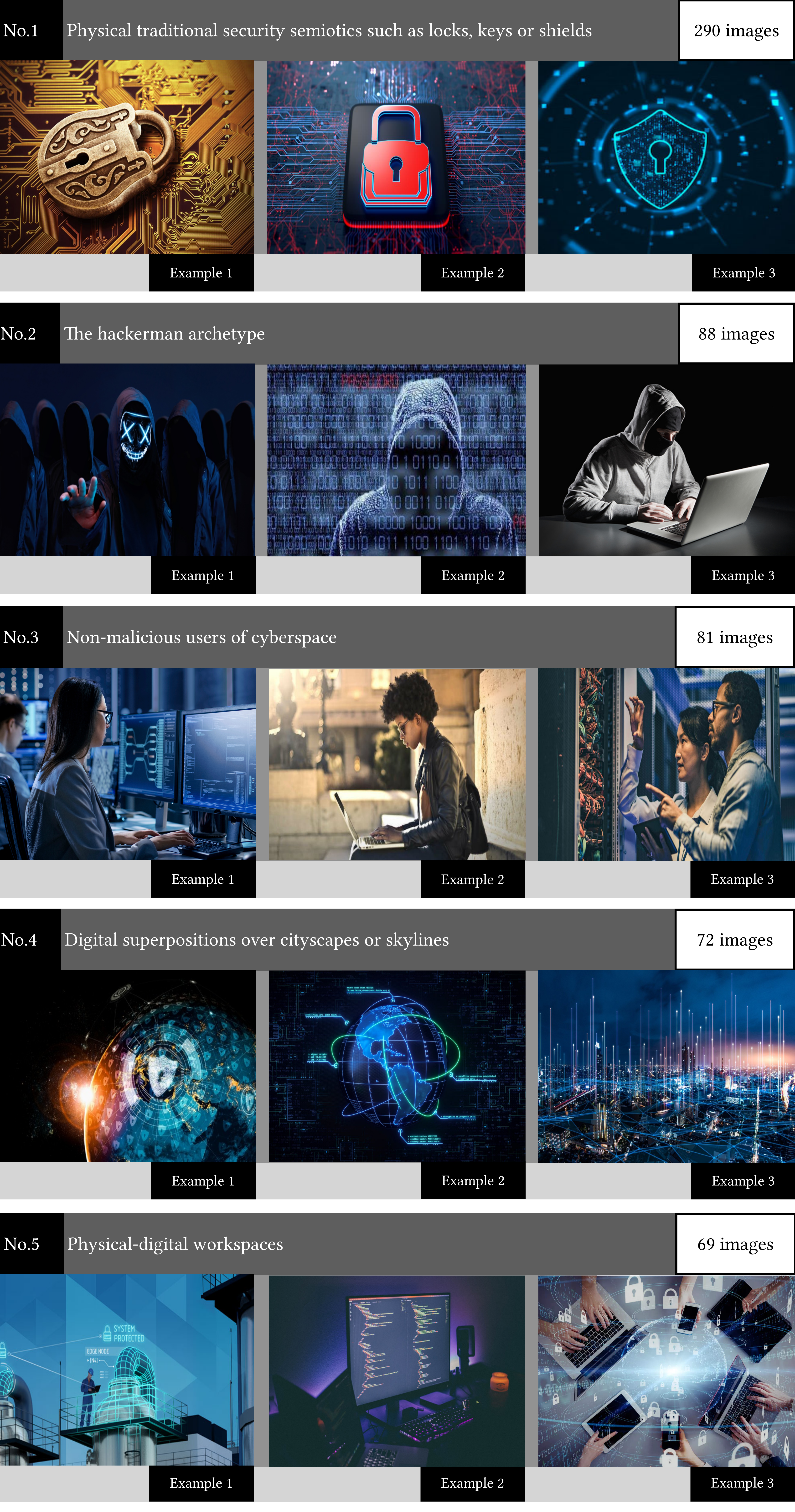}
    \caption{A random selection of images taken from classes 1-5.}
    \label{fig:classes1-5}
\end{figure}

\begin{figure}
    \centering
    \includegraphics[scale=0.118]{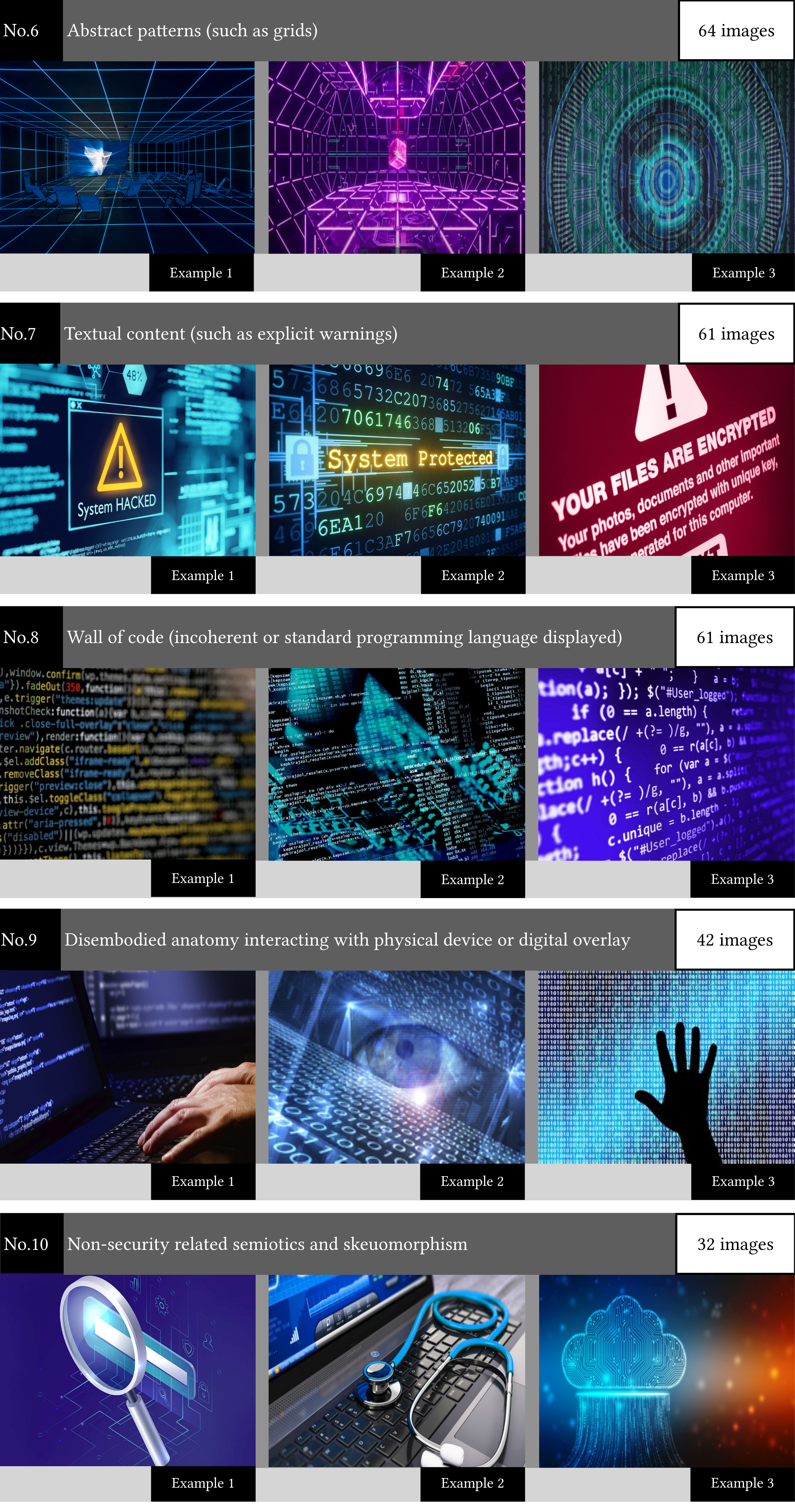}
    \caption{A random selection of images taken from classes 6-10.}
    \label{fig:class6-10}
\end{figure}

Using the process described in Sections~\ref{search} and \ref{cleaning}, we compiled an image-set that covered a wide swathe of cyber security topics and their associated aesthetics.  Small selections of images from each class are shown in Figures~\ref{fig:classes1-5} and \ref{fig:class6-10}.  Through the process described in Sections \ref{sec:data} and \ref{colour}, we identified 32 distinct and internally consistent image classes in the image-set.  These ranged from abstract interpretations of networked security to imagery depicting the binary view of cyber security as an eternal battle between malicious actors and their victims.  However, because most of the images (80.6\%) were concentrated in just ten major classes (as detailed in Table \ref{tab:classes}), we restrict our discussion to these classes.  We further group these classes into four broad (but not mutually exclusive) categories: Objects, People, Places, and Others.

\begin{table}[t]
    \centering
       \caption{The top ten classes in our image-set.}
    \begin{tabular}{|p{0.75cm}|p{5.4cm}|p{0.5cm}|}
    %\begin{tabular}{l|l|c}
\hline
\textbf{Class} & \textbf{Description} & \textbf{No.}\\
\hline
1. & Physical traditional security semiotics (such as lock, key, or shield) & 290\\
\hline
2. & Hackerman archetype & 88\\
\hline
3. & Non-malicious users of cyberspace & 81\\
\hline
4. & Digital superpositions over cityscapes or skylines & 72\\
\hline
5. & Physical-digital hybrid workspaces & 69\\
\hline
6. & Abstract patterns (such as grids) & 64\\
\hline
7. & Textual content (such as explicit warnings) & 61\\
\hline
8. & Wall of code (incoherent or standard programming language) & 61\\
\hline
9. & Disembodied anatomy interacting with a physical device or digital overlay & 42\\
\hline
10. & Non-security-related skeuomorphism & 32\\
\hline
\end{tabular}
 
    \label{tab:classes}
\end{table}

\subsection{Colour}

As one would expect, colour features heavily across all of the classes in our image-set. If we use the heat-maps to discuss similarity potential across colour and texture (where radical differences may indicate domain-shift within the class \cite{samek2016evaluating}), then this section concerns itself with the qualitative \textit{use} of colour as we see it across the complete image-set. 

Where many of the abstract forms seen in Class 6 utilise hues running the gamut from greens to blues which are almost always contrasted by dark backgrounds, it is clear from the complete image-set that no `universal' definition or convention on the usage of colour exists within cyber security, beyond the heavy use of cyan blue\footnote{This was already informally known to the cyber security community, where the cyan blue code \#235594 was the most commonly found colour in a large image-set, but which is not explicitly referenced to here due to the fact it was not cleaned for duplicates and other issues, and contained little in the way of search methodology. It can be found here: \url{https://daylight.berkeley.edu/cybersecurity-imagery/}.}. 

All of the heat-maps highlight these similarities in colour. Where we find points of interest are in Class-1, where colour is used to denote objects as being of specific importance, ranging from useful to dangerous. We may contrast this with another aspect of cyberspace, the video game, where emphasis is placed on the colour of objects with which the player may interact~\cite{johnson2017history}. This codification of objects again lacks a specific narrative, and objects that are beneficial often share hues with those semiotics deemed dangerous.

\begin{figure}[t]
    \centering
    \includegraphics[scale=0.175]{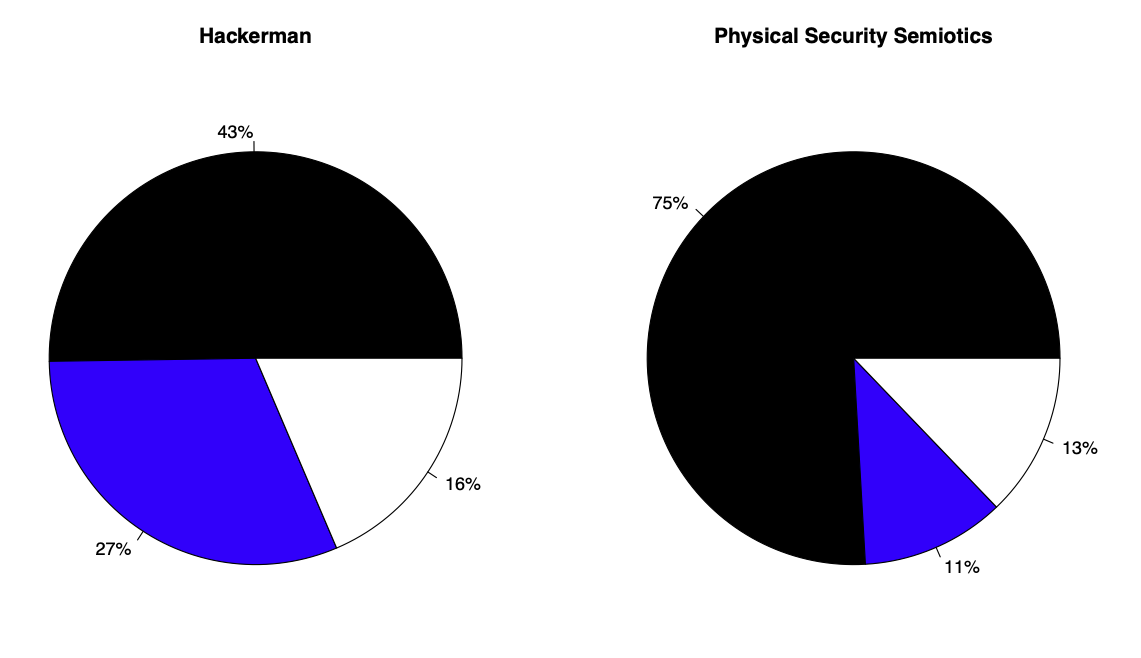}
    \caption{An example of the most commonly used colours within our top-two classes, simplified to 4-bit.}
    \label{fig:colouranalysis}
\end{figure}

In between the classes, a specific colour analysis leads to results of limited immediate utility. Turning our attention to Figure~\ref{fig:colouranalysis}, we see that our top-two image classes share a penchant for blacks, blues and whites. 

With this large amount of blacks and darker hues serving as background, we see in many classes a contrasting effect between the brightly coloured objects, spaces and vertices. In this manner, we hypothesise that colour may be used to draw the user's attention to these objects, which exist in a space with no other domains to draw inspiration from, similar to video games~\cite{johnson2017history}. A colour analysis may not allow us to infer further specifics as to their self-efficacy potential, in which the colours may be used to deliver a decoding of the messages present in the accompanying textual content. Instead, we look towards the objects, people, places and other aspects of the image. In these cases, we find the heat-maps to be of more use as an accompaniment to assess the credibility of the assertions. Each heat-map represents the relationship any given image has to the other images within its class, with blue cell colours indicating greater similarity and red cell colours indicating lesser similarity.

\subsection{Objects}

\begin{figure}[t]
    \centering
    \includegraphics[scale=0.2]{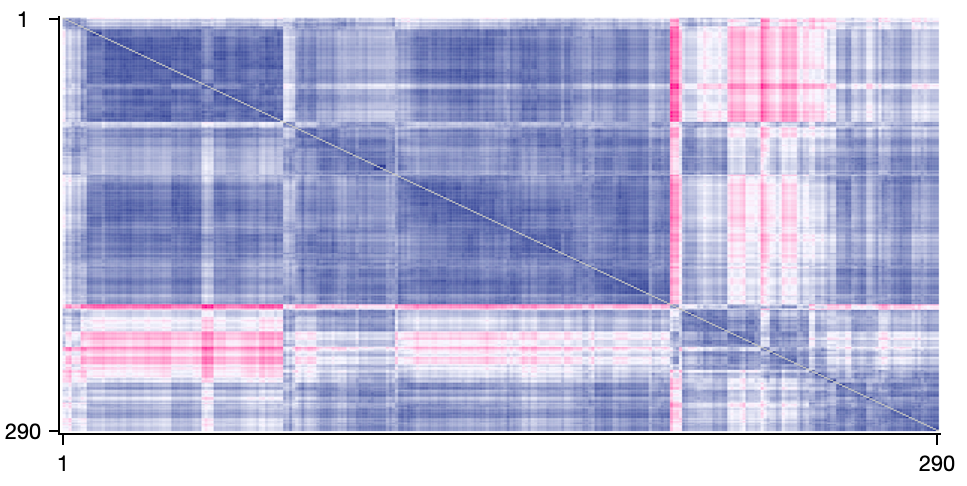}
    \caption{Heat-map highlighting the overall colour differences between images in Class 1 (Traditional physical-digital security semiotics).}
    \label{fig:class1heatmap}
\end{figure}

\begin{figure}[t]
    \centering
    \includegraphics[scale=0.2]{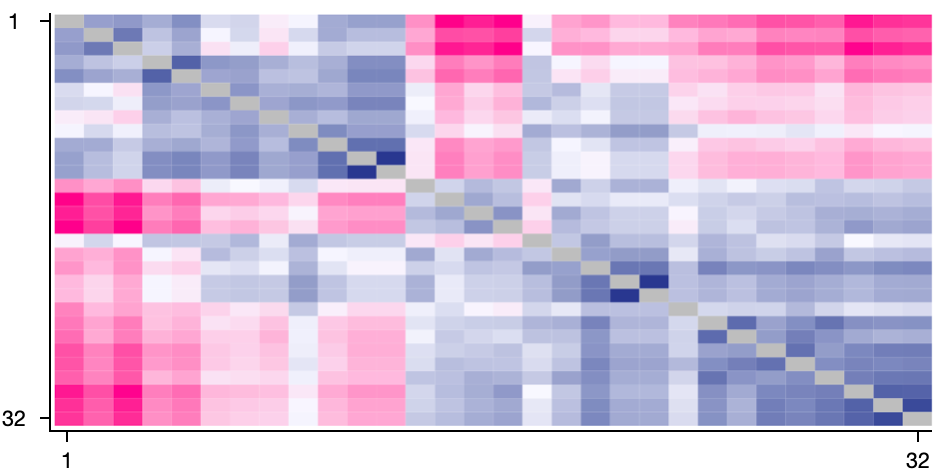}
    \caption{Heat-map highlighting the overall colour differences between images in Class 10 (Non-security semiotics and skeuomorphism).}
    \label{fig:class10heatmap}
\end{figure}

Classes 1 and 10 feature objects associated with physical security, like locks, keys, and shields (and others identified by \cite{ghertner2020futureproof}); objects appropriated from non-security domains, such as cameras; and skeuomorphic adaptations of real-world objects, like digitised versions of envelopes.  Class 1 contains 290 unique images with extraordinary colour similarity (as per Figure \ref{fig:class1heatmap}), implying consistent use of similar semiotics and colour schemes throughout.  %It bears repeating that these represent unique images.
Class 10 is much smaller, featuring only 32 unique images, but it is notable for its wider variety of objects and colours (see Figure \ref{fig:class10heatmap}).  Figure \ref{fig:class6-10} highlights some of the images from these classes.

\subsection{People}

\begin{figure}[t]
    \centering
    \includegraphics[scale=0.2]{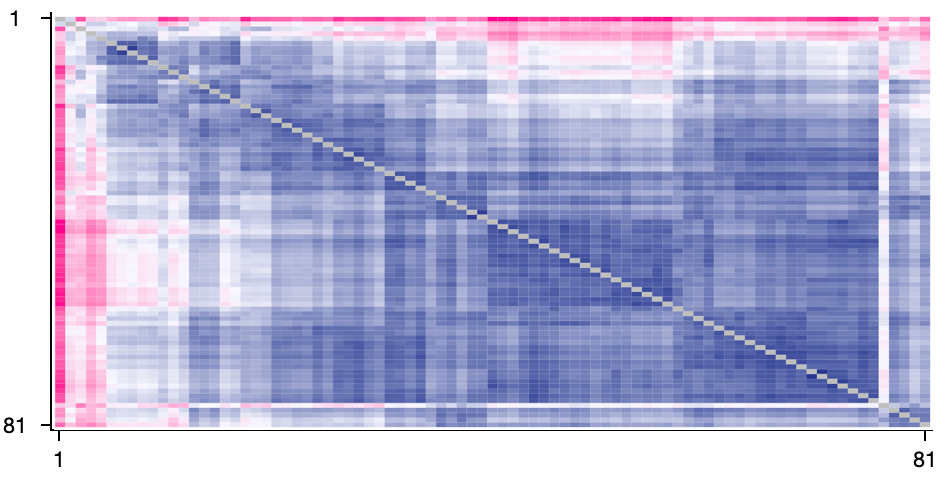}
    \caption{Heat-map highlighting the overall colour differences between images in Class 2 (Hackerman archetype).}
    \label{fig:class2heatmap}
\end{figure}

\begin{figure}[t]
    \centering
    \includegraphics[scale=0.2]{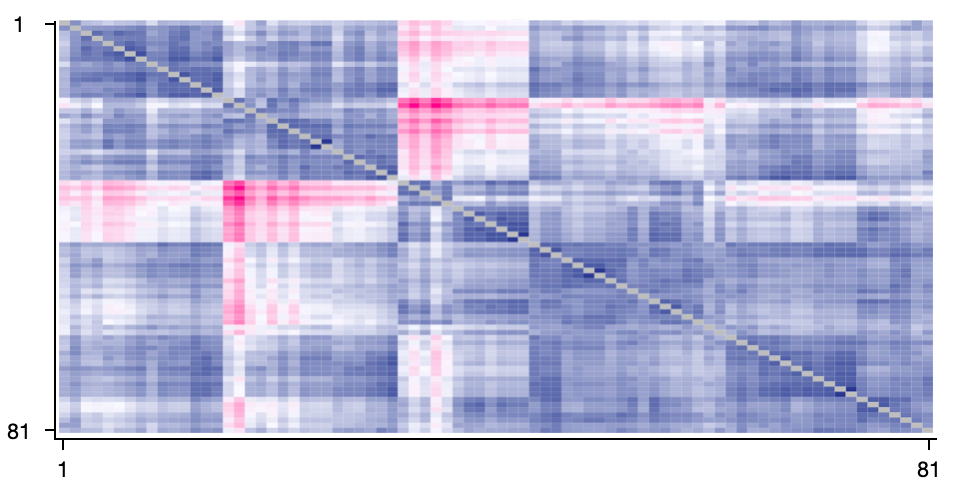}
    \caption{Heat-map highlighting the overall colour differences between images in Class 3 (Non-malicious users of cyberspace).}
    \label{fig:class3heatmap}
\end{figure}

\begin{figure}[t]
    \centering
    \includegraphics[scale=0.2]{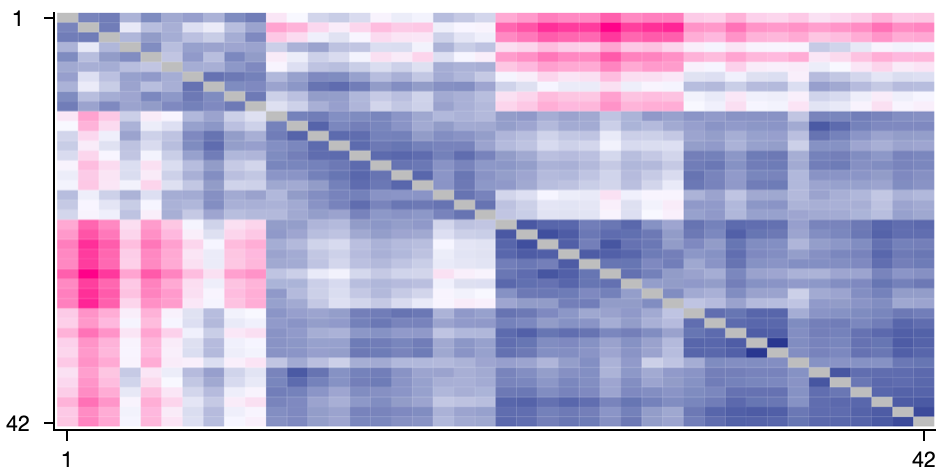}
    \caption{Heat-map highlighting the overall colour differences between images in Class 9 (Disembodied anatomy interacting with a physical device or digital overlay).}
    \label{fig:class9heatmap}
\end{figure}

Classes 2 and 3 feature individuals who are implied to be malicious (Class 2) or non-malicious (Class 3) users of cyberspace.  We also include Class 9 within this group, given its emphasis on human anatomy.  Class 2 contains 88 unique images with significant colour similarity (see Figure \ref{fig:class2heatmap}), which in this case implies similar compositions~---~individuals assuming similar stances against similar (dark) background colours.  Class 3 consists of 81 unique images and is slightly more varied in its make-up (as per Figure \ref{fig:class3heatmap}).  Class 9 is the smallest and most differentiated class in this group, consisting of only 42 unique images with wider colour discrepancies in the heat-map (see Figure \ref{fig:class9heatmap}).  Figure \ref{fig:classes1-5} highlights some images from Classes 2, 3, and 9 are shown in Figure \ref{fig:class6-10}.

\subsection{Places}

\begin{figure}[t]
    \centering
    \includegraphics[scale=0.2]{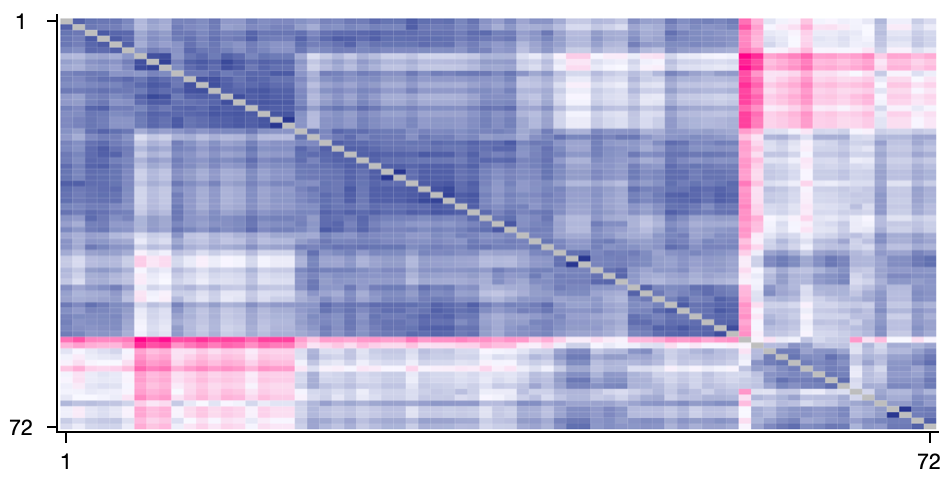}
    \caption{Heat-map highlighting the overall colour differences between images in Class 4 (Digital superpositions over cityscapes or skylines).}
    \label{fig:class4heatmap}
\end{figure}

\begin{figure}[t]
    \centering
    \includegraphics[scale=0.2]{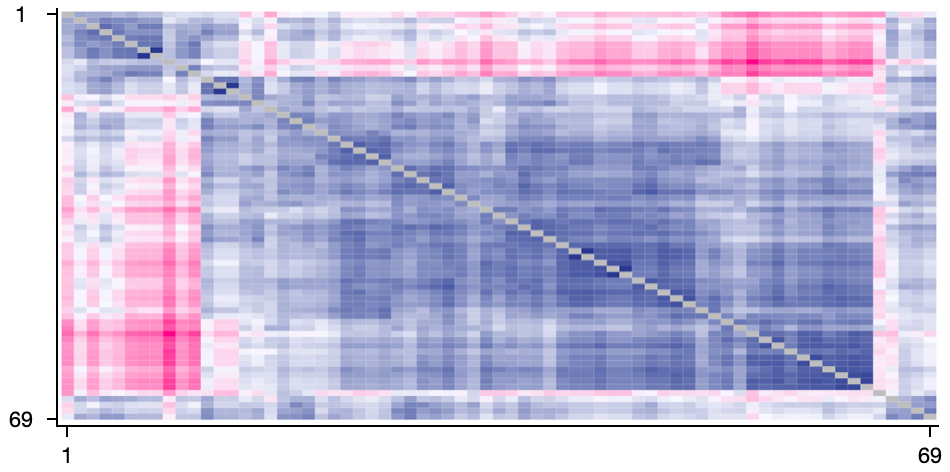}
    \caption{Heat-map highlighting the overall colour differences between images in Class 5 (Physical-digital hybrid workspaces).}
    \label{fig:class5heatmap}
\end{figure}

Classes 4 and 5 feature specific places related to cyberspace and cyber security, such as futuristic urban spaces (Class 4) and workspaces (Class 5).  Class 4 contains 72 unique images with significant colour similarity (see Figure \ref{fig:class4heatmap}), once again implying similar compositions and colour palettes.  Class 5 consists of 69 unique images and is more varied, as can be seen in Figure \ref{fig:class5heatmap}.  Figure \ref{fig:classes1-5} highlights some examples from Classes 4 and 5.

\subsection{Other}

\begin{figure}[t]
    \centering
    \includegraphics[scale=0.2]{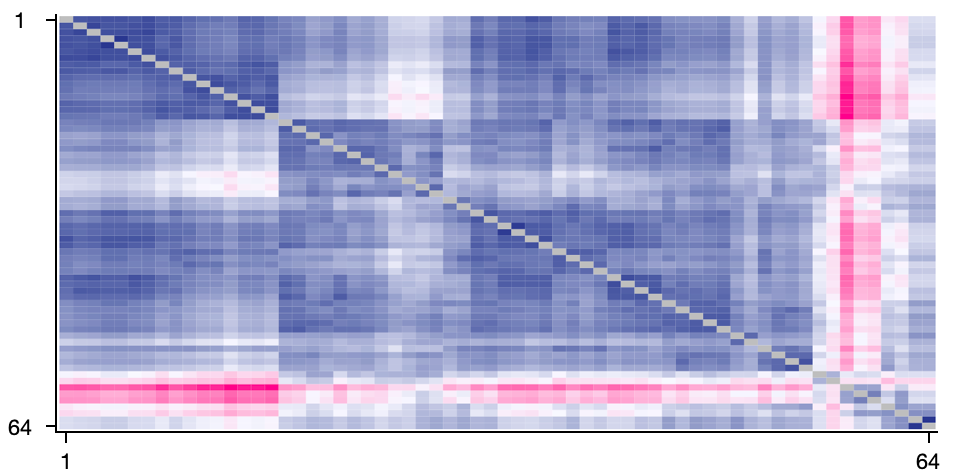}
    \caption{Heat-map highlighting the overall colour differences between images in Class 6 (Abstract patterns).}
    \label{fig:class6heatmap}
\end{figure}

\begin{figure}[t]
    \centering
    \includegraphics[scale=0.2]{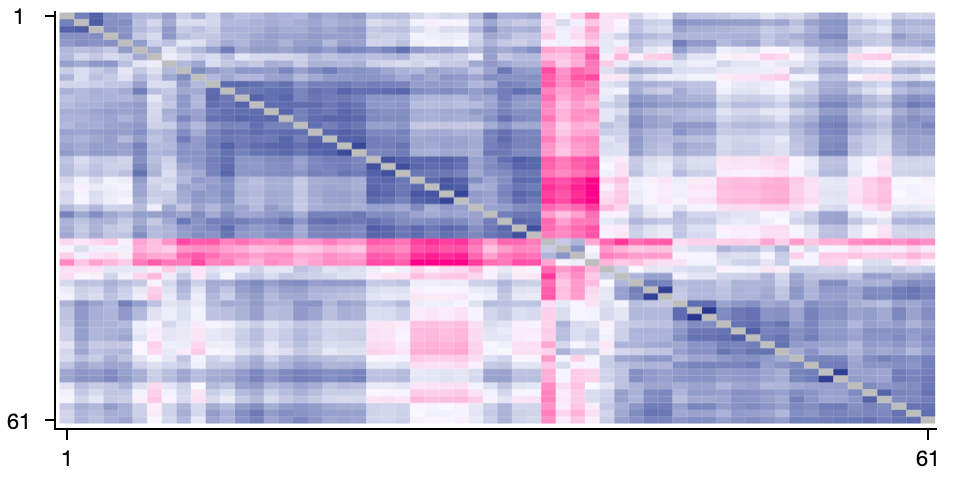}
    \caption{Heat-map highlighting the overall colour differences between images in Class 7 (Textual content such as explicit warnings).}
    \label{fig:class7heatmap}
\end{figure}

\begin{figure}[t]
    \centering
    \includegraphics[scale=0.2]{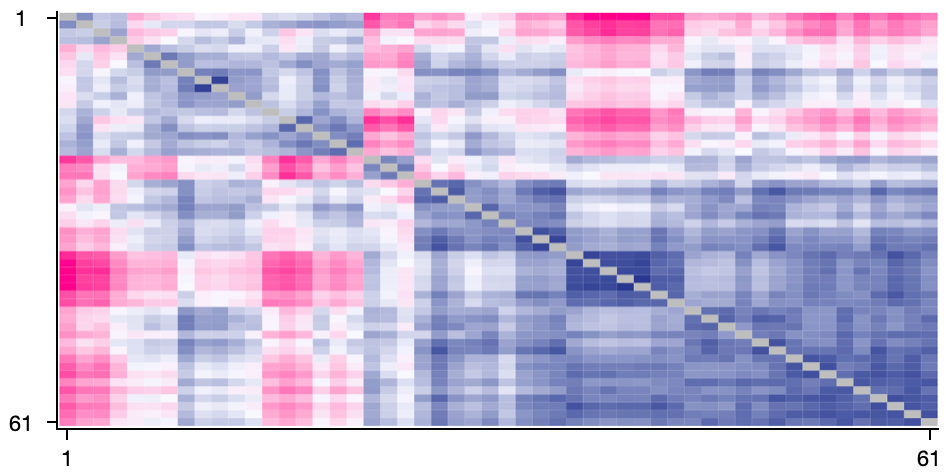}
    \caption{Heat-map highlighting the overall colour differences between images in Class 8 (Wall of code).}
    \label{fig:class8heatmap}
\end{figure}

Classes 6, 7, and 8 variously encompass imagery of digital patterns, alphanumeric symbols, and other two-dimensional or abstract representations of cyberspace and cyber security.  Class 6 contains 64 unique images with reasonable colour similarity (as per Figure \ref{fig:class6heatmap}); though similar background colours are frequently used to represent mathematically defined patterns and shapes, there is some variation based on other, supporting semiotic attributes in this category.  Class 7 consists of 61 images that are slightly more varied, as seen in Figure \ref{fig:class7heatmap}.  Meanwhile, Class 8 contains 61 images, and it is the most diverse of the three (as visible in Figure \ref{fig:class8heatmap}.  However, the representativeness of Class 8's heat-map is limited by the content of the images, namely incoherent alphanumerical symbols or programming languages on a dark background.  We expect that the variation in the heat-map reflects the wide variety of colours used in these different symbols, despite larger compositional similarities.  Figure \ref{fig:class6-10} highlights some examples from Classes 6, 7, and 8.

\section{Discussion} \label{Discussion}

In this section we argue that the aesthetic classes established in the previous section can help to prime readers' interpretations of associated texts in ways that affect their self-efficacy.  
%In order to answer it sufficiently, we begin by looking at the semantics of cyber security aesthetics, that is to say, constructs with universally accepted meanings, such as basic shapes and colours. We follow these up by analysing our top ten image classes, and associate these with specific elements that may harness self-efficacy potential.
Insofar as the images that accompany cyber security texts are created and selected for this specific purpose (that is, for inclusion in these texts), we may say that the observed aesthetic trends are more or less deliberate attempts to frame cyber security in a certain manner.
Accordingly, we begin with an analysis of the semantics of cyber security~---~the colours, shapes, and devices used throughout the image-set~---~to understand how cyber security is being framed.  We proceed to analyse the image classes that feature objects, people, and places.  We finish this section by framing these discussions in light of our research aims and what we learnt in Section~\ref{Background}.

\subsection{The visual vocabulary of cyber security}

Our image-set suggests that cyber security exists at the limits of traditional human visibility.  Indeed, many of the image classes feature abstractions of objects, situations, individuals, and landscapes rather than concrete subjects.  This seems consistent with typical evocations of `cyberspace'~---~a term coined in 1982 by science-fiction writer William Gibson~\cite{gibson1982neuromancer} to designate `a new universe' parallel to the physical but created by the digital~\cite{benedikt1991cyberspace} (as encapsulated in the `fifth branch' metaphor utilised by the U.S. military~\cite{branch2020s}).  Although the term has since become synonymous with global computer networks such as the Internet, it continues to encapsulate the `sublime' sensations associated with a new frontier.  \cite{nye1996american} explained that, when users are introduced to powerful new technologies (such as cyberspace or a digital system), they experience a pleasurable yet terrifying sensation that alerts them to the limits of their reality.

Multiple scholars have explored how aesthetic choices can help to acclimate users to the new reality of cyberspace `environments'.  \cite{featherstone1996cyberspace}, for instance, argued that aesthetics help to evoke imagery of life in the domain of cyberspace, while \cite{croon1999making} explained that imagery can provide insight into a domain and help to cultivate a form of spatial awareness within it.  Within our image-set, we observed heavy use of mathematical aesthetics such as concentric arcs, simple and tileable shapes (such as hexagons, albeit with a higher area-to-perimeter ratio), and connecting lines.  Where these devices construct the \textit{shape} of cyberspace, \textit{colour} establishes its tone.  In our image-set, we observed a preponderance of dark shades and blue hues. This is consistent with the work of \cite{shedroff2012make}, who researched the use of colour within the domain of science fiction.  One possible reason for this preference in science fiction and cyber security is the relative rarity of blues in nature~\cite{greenspan_2013}; because this colour is scarce in our physical domain, it effectively communicates cyberspace's distinction from the straightforwardly physical and natural.

Of the aesthetic classes identified in our image-set, the ones that most optimally represent cyberspace as a techno-spatial domain separate from, but parallel to, our physical reality are Classes 3, 6 and 8, which represent 16\% of our total image-set.  These classes deploy the aforementioned aesthetic tactics most consistently / legibly.

\subsection{Physical traditional security semiotics}

%\begin{figure}[t]
%    \centering
%    \includegraphics[scale=0.2]{Assets/semioticsample.png}
%    \caption{A selection of the semiotics used for aesthetic or narrative purposes.}
%    \label{fig:semiotics}
%\end{figure}

\cite{beasley2010persuasive} argue that semiotics researchers are mainly interested in understanding how individual signs, objects, and concepts coalesce into a coherent visual narrative.  Objects such as locks, shields, and keys appeared frequently in our image-set, either as digital manifestations of physical objects or, in rare cases, as physical objects in a hybridised physical-digital environment.  These objects represent 20\% of the total image-set, and a selection can be found in Figure~\ref{fig:classes1-5}.

%In many cases however, our image-set contains these objects in such a manner that the overall narrative may be obscured.

As these objects often pre-date the development of cyberspace, they may initially evoke physical, rather than digital, security, being grafted onto cyber security only later to represent what security might mean in cyberspace.  As is the case with skeuomorphic interface design, wherein new signs or symbols are developed from prior objects (retaining the original's ornamental design cues), the new semiotic interpretation may have superseded the older meaning.  
%Semiotics within interface design have tended towards the skeuomorphic, in which new signs or symbols are developed out of prior objects, retaining the original's ornamental design cues, in order to aid the viewer in understanding.  It is possible for the new object to supersede the original, becoming the \textit{de-facto} natural symbol for said domain.
Nonetheless, any theoretical constructs associated with the originator sign have likely been carried forward~\cite{kearney2001continental}.

According to \cite{ghertner2020futureproof}, security could be seen as a form of negation, with security signs suggesting the \textit{absence} of malicious activity.  %The manner of which this absence is manifested within cyber security visuals can most often be seen in the form of semiotics.  In our case, we could conceive of all these objects providing negation to provide a form of security.
The lock is a powerful symbol of security in the real world, the key a symbol of legitimate authority, and the shield a symbol indicating defense in the event of an attack.  These symbols can also be rendered differently to present a kind of advanced warning system; imagery of a broken lock, key, or shield might alert the user to potential security infractions.\footnote{It must be pointed out, however, that we did not come across any broken or damaged semiotic objects in our image-set.}

\subsection{Hackers: anatomy, gender, and race}

Humans were represented in 16.4\% of the images in our image-set.  Using our codebook, we assigned each of these images to one of two binary categories: the malicious \textit{hackerman archetype} (8\% of our total image-set, and just over 52\% of our total human representations) or non-malicious users (7.8\% of our total image-set, and just under 48\% of our total human representations).  An additional 4\% of our image-set contained anatomical images, such as those seen in Class 9.  These anatomical images were largely restricted to two types: those featuring hands and those featuring eyes.

Our image-set broadly suggests a lack of equal representation in cyber security aesthetics, and thus a lack of diverse role models.  Just under 1\% of our hackerman archetype features feminine-presenting people; all others were either masculine-presenting (63\%), implied to be men, or had their secondary-sex characteristics obscured (for instance, by a mask).  While we do see a wider variety of individuals amongst the non-malicious users~---~63\% of these images contained masculine-presenting individuals, 37\% contained feminine-presenting individuals, and 20\% contained people who appeared to be of a non-white background~---~this does not necessarily represent the gender or racial/ethnic background ratio of audiences who engage with cyber security content, but that of the organisations that produce cyber security digital media
(as per Thomas's \cite{thomas2020discursive} concept of a `discursive digital archive').\footnote{Statistics from the U.S. Bureau of Labor Statistics indicate that 18\% of cyber security specialists are women~\cite{jethwani2017can}, which may provide a reference point for the 37\% of human images featuring feminine-presenting people.}

Our image-set also seems to attribute disparate expertise~---~and thus responsibility for cyber security~---~to different individuals.  The hackerman, for instance, is presented as a `lone wolf', whereas non-malicious individuals are more frequently depicted in groups than as solo actors.\footnote{This is particularly pronounced in images featuring feminine-presenting people, who in many cases are accompanied by masculine-presenting people.}  One could interpret this to mean that individuals lack the talents or other requisite knowledge that the lone hackerman possesses~\cite{hack}, and must therefore work together to counter the hackerman's threats.  The media promotes this association by applying the hackerman stereotype to organisations such as `Anonymous' and online communities such as 4chan, placing these groups in a position to dominate the conversation. In turn, individual users may view these entities as malicious experts, abdicating their own responsibility for cyber security based on their feelings of powerlessness~\cite{hack}. 

Finally, our image-set features moral ambiguity and vague representations that could make it difficult for users to derive context from images.  For example, while the hackerman is supposed to engage in malicious online activities, he is sometimes presented with morally ambiguous or vigilante imagery like the `V for Vendetta' mask, complicating the viewer's understanding of his aims.  While non-malicious individuals are consistently depicted as benign or neutral, they are frequently engaged in a variety of nondescript tasks, like interacting with hacked devices, responding to being hacked, or performing some kind of professional work (for instance, as cyber security professionals).  Images of human anatomy were similar; though hands are an important and highly visible part of the human body~\cite{jakubietz2005defining} that can serve as a heuristic to facilitate learning~\cite{goldin2005our},\footnote{Hands are often used to model movements or convey information by assuming specific positions.} in our image-set they were engaged in a variety of mostly unclear / un-directed movements and positions, often in connection with physical devices like laptops.  Images in which the eye was dominant were similarly varied, but in many cases represented the moral ambiguity of a panopticon (alongside images from other classes that render a retina and cornea from composite imagery) or enjoined the user to pay attention~\cite{gaines2019machinic}.

By depriving users of role models, clear contexts / goals, and the means or abilities to achieve such goals, these issues likely undermine individual users' self-efficacy.  We believe that these issues stem, at least in part, from cyber security's reliance on stock photography.  Stock photography is characteristically nondescript and visually homogeneous because it must make individual images salient to various use cases~---~hence the images of individuals in vague contexts and unclear narratives.  Furthermore, stock photography in western media features a `discrimination implied by a well-calculated, almost mandatory inclusion of gender and ethnic minorities'~\cite{papadopoulou2014seen}, yielding the nominally inclusive images that nonetheless fail to actually bestow agency to the individuals represented.

\subsection{Digital-Physical spheres}

Representing 7\% of our total image-set, hybrid digital-physical representations, such as digital networks superimposed over cityscapes, the earth, or a more abstract sphere, were of some note to us.  To understand what these images mean and how they may affect self-efficacy, we consulted work in other domains that use similar styles of visualisation.  Most prominently, Sloterdijk, a German philosopher, studied the history of spherical maps, overlays, and designs, tracing these visual tactics as far back as the late 15th Century~\cite{sloterdijk2011spheres}.  Sloterdijk argues that these kinds of images arose naturally from our changing understanding of our planet at that time, which was no longer an enclosed space or the centre of the universe, but a single, contingent celestial body.  According to this account, disoriented European map makers began to fetishise spherical imagery as a sense-making device capable of conveying that we were no longer living inside a world, but rather \textit{on} one.

Although the overlays in our image-set differ from these earlier precedents in that they overlay networked security graphs instead of shipping lanes, they may nonetheless embody a return to such `spheric-security' in the face of new metaphysical uncertainties, at least from an aesthetic viewpoint~\cite{sloterdijk2011spheres}.  Indeed, a subconscious spheric-security can also be seen in images outside our image-set like network traffic maps, which often take a spherical view despite not being technically constrained in this manner.  According to \cite{10.1093/cybsec/tyv004}, spherical shapes suggest something which can be contained and kept secure, and so they may help to imaginatively `bound' the sprawling endlessness of Network traffic, which always threatens to deviate into the unknown.

\subsection{Trends and Recommendations}

In the above, we explored the attributes of cyber security aesthetics and speculated about the kind of self-efficacy that these attributes afford.  In the following, we highlight the trends we observed across classes and the key issues that cyber security aesthetics must address to improve its effect on users' self-efficacy.

\subsubsection{A practical philosophy for cyber security}

According to the philosophical discourse on aesthetics presented in Section~\ref{Background}, aesthetics provide a basis for savoring sensation, organising sensations into orderly meaning~\cite{Kant1892-KANTCO-3}, and orienting contexts~\cite{kikuchi1992philosophic}.  Moreover, the meanings and contexts thus constructed can influence our decisions~\cite{chatterjee2014aesthetic}.  While cyber security aesthetics seem to fulfil the first element of sensation, adapting a long history of mathematical and even science fiction aesthetics (alongside other disciplinary symbols) to frame cyberspace as an other-worldly future domain that is not restricted to what we currently understand, 
%What we find of particular interest is the way in which these images may have an effect on self-efficacy.
they fail to provide sufficient context for effective navigation or learning.  
%We understand the purpose of these images is to aid the learning process (as discussed in Section \ref{Background}), however we do not believe the image-set fulfills this mandate.
Instead, they present a dazzling spectacle of abstractions and powerful traditional security semiotics without much in the way of meaning.  
%The image-set represents a kind of convenience, an elaborate spectacle with little substance or meaning.
The hackerman archetype, powerful but elusive, and our other subjects involved in procedural and ambiguous work, encourage users to abdicate responsibility based on their own comparative lack of expertise and perceived powerlessness, while the digital-physical spheres, like Sloterdijk's historical spheres, reify a world that may not match our experience of reality.  According to Sloterdijk, humans no longer believe in an all-seeing singularity encompassing us, be it supernatural or human exceptionalism~\cite{sloterdijk2011spheres}. Nonetheless, much user-facing cyber security media depicts cyberspace as a fraught and hostile environment that individuals can't hope to navigate without expert assistance, a strategy that amounts to `fear-mongering' and allows those with vested interests in the cyber security field to turn security into an all-important and all-encompassing issue~\cite{neocleous2008critique}.

Solving these problems is no simple task, and so no easy solution presents itself.  One high-level, long-term suggestion is to develop a practical philosophy based on narratives.  For example, \cite{mcsweeney1999security} frame security as a form of resilience that individuals can build through everyday tasks that make them feel secure.  Where the person performing these tasks is a trusted friend or confidante, security could also be framed as \textit{relational} security.  Breaking cyber security down into small, actionable tasks can significantly improve users' self-efficacy, given the theory of self-efficacy presented in Section~\ref{Background}.  It would also allow cyber security aesthetics to play a positive role, enabling individuals and society at large to navigate the rapidly changing digital-physical world represented in our image-set.  This possibility could be realised through new and innovative semiotics that are not simply copies of the traditional security landscape, or through a more positivist and standardised form of low-abstracted relational imagery with clear links to cyberspace. A practical example of this can be found in refuse recycling, where Gary Anderson, a student, won a nationwide contest for a new symbol for the then fledgling recycling initiative with his Mobius loop-based three-chasing-arrows. This symbol has since risen to global prominence \cite{jones1999gary}. 

\subsubsection{Improving role models for cyber security}

Systemic under-representation in certain occupations is a complex and multi-causal problem that needs to be examined using both interdisciplinary and context-specific approaches.\footnote{These approaches must also account for under-representation at multiple phases / points in time, including factors that influence the admission, participation, and progression of under-represented individuals in these industries.}  Fortunately, we can begin to bolster all users' self-efficacy through much more straightforward steps.\footnote{In our case, self-efficacy is linked to the context in which these images are used for informally learnt cyber security.}  For example, insofar as stereotypes can have positive self-efficacy effects in certain contexts~\cite{czopp2015positive},\footnote{In the field of education, positive stereotypes have been shown to influence which goals individuals choose to pursue~\cite{czopp2015positive}.} we could co-opt the male-dominated hackerman stereotype and make it more inclusive, extending its connotations of moral ambiguity and power to individuals across demographics.  We believe that such role models can help to reduce users' cognitive load when assimilating cyber security knowledge, and that they can make the field as a whole seem more user-friendly.
 
\subsubsection{The paradox of simplification}

In the world of user interface design, developers build persuasive and easy-to-use interfaces to pursue a kind of universal simplicity.  This objective could also be applied to cyber security aesthetics, where it could help to magnify users' self-efficacy.  For instance, cyber security aesthetics could become simpler, cleaner, and more pertinent to the subject matter, or they could feature more rhetorical imagery.  Where the digital workspaces, networked landscapes, text walls, and semiotic padlocks in our image-set all pose exploratory questions (prompting internal narrative reflection), rhetorical imagery makes a specific point, is designed with a specific audience in mind, and is focused on narrative integrity above all.

Of course, the problem is not just about exploratory imagery; we have seen that cyber security aesthetics feature many abstractions and visual devices that confuse the core concept being depicted.  If cyber security aesthetics are supposed to simplify the complex nature of cyber security, it appears that we must first simplify the aesthetics themselves.  However, simplification can itself lead to misunderstanding, as was the case in the Space Shuttle disaster, which was partially attributable to the oversimplification of data in a graph~\cite{tufte2006beautiful}.  This is the cyber security aesthetic paradox: that simplification can aid as well as hinder understanding in equal measure.  To overcome this challenge, we might look to other fields that have developed unique aesthetic norms that enhance learners' self-efficacy.  The field of chemistry, for instance, has spent centuries developing a standard aesthetic system that simplifies complex concepts and narratives without rendering them ineffective~\cite{hoffmann2003thoughts}.

\section{Limitations and future research directions} \label{future}

The scope of this study was limited by the definitions we used and the selection criteria we applied to guide our assembly of the image-set.  In our case, this meant focusing on English-language material even though a preliminary search conducted before implementation unearthed a rich catalogue of images in other languages.  This also means that our analysis and our findings likely exhibit Anglo-Saxon bias.  Nevertheless, we expect that our methodology can be adapted to explore the same research aims in other languages and cultural contexts, enabling a more universal understanding of cyber security aesthetics.

Given the background we provided in Section \ref{Background}, our definition of aesthetics likely exhibits Anglo-Saxon or broadly Euro-American bias, failing to encapsulate the aesthetic philosophies of other cultures. It may also be limiting for other reasons, as we focused narrowly on their ability to provide context for, and inlays to, the world of cyberspace, making it more interpretable and navigable for informal learners.  Other definitions might yield different insights and support other kinds of research questions.

We utilised semi-automated methodologies to classify images based on the semiotic objects within them, and the results are tempered by the respective limitations of these methodologies.  Moreover, our results represent a specific snapshot in the security timeline; access to a larger historical image-set would inevitably change the overall results, potentially yielding a more statistically significant sentiment analysis.  Furthermore, in order to assemble a unique image-set, we ignored duplicates.  However, insofar as aesthetic literacies arise from \textit{exposure} to images, rather than the absolute number of unique images, including duplicates could help us to ascertain the effective rate of gender representation, as perceived by viewers.

This study was exploratory, and we predicted how users' self-efficacy might be affected by images through a singular focus on qualitative analysis.  We did not consider other metrics that could have enhanced the findings, and we did not engage in other kinds of data collection (like surveys) that could have revealed real users' reactions.  Traditionally speaking, image recognition assessments in lab settings involve comprehension tests, eye tracking, and brain-imaging.  Knowledge of how users interact with cyber security aesthetics in these terms would allow for a significantly richer analysis of aesthetic effects on self-efficacy.

Finally, because self-efficacy is a fluid construct that may vary based on specific emotional awareness and specific tasks or contexts~\cite{karademas2007optimism}, future research in this field could assess the emotional associations of more granular aesthetic elements, like each colour within each image class.  For example, \cite{mohammad2013colourful} crowd-sourced an inventory of colour-word associations that reveal the specific emotions attached to specific colours, which could be used to analyse the colour trends in our image-set.

All of these limitations present myriad opportunities to expand on this work, perfecting our understanding of cyber security aesthetics and the effects it can have on cyberspace and its users.

%narrowly on the role of images in international politics, rather than the visual world generally (e.g., Andersen, Vuori, & Guillaume, 2015; Heck & Schlag, 2013; Methmann, 2014).

\section{Conclusion} \label{Conclusion}

In this paper we have presented work on an image-set of cyber security aesthetics generated from mainstream media articles as might be faced by individual users on a regular basis. The work was oriented by two aims: (1) to ascertain what cyber security aesthetics consist of, from the perspective of an individual user, and (2) to provide an explorative discussion as to the manner in which these aesthetics may affect users' perceived self-efficacy as they informally learn cyber security precepts.

Our findings for (1) indicate that cyber security aesthetics depict a threatening and confusing environment with systemic semiotic and social deficiencies~---~a distorted vision of cyberspace without clarity of thought. The narrative of cyber security is abstract and opaque, but through informal learning, individual users can assemble a mental representation of this concept, using their senses to intuit meaning and perspective from the visual elements that accompany cyber security texts.  

For (2) our findings raise important obstacles to self-efficacy potential, from the way that participants of cyberspace are portrayed to the moral ambiguity that characterises a significant proportion of the image-set.  Nonetheless, as cyber security continues to evolve into a core concept within cyberspace, we believe that these issues can be overcome. Indeed, several of these problems, and the simplification paradox itself, arise from a lack of vision, or perhaps just the lack of usable semiotics available to cyber security content creators. We believe the work represents the first steps in working towards a more holistic and cohesive cyber security aesthetic vision.

\backmatter

\noindent

\bigskip

%\begin{appendices}

%\section{Section title of first appendix}\label{secA1}




%\end{appendices}




\section*{Statements and Declarations}

All authors contributed to the study conception and design. Material preparation and data collection were performed by [Mark Quinlan]. Initial data analysis of the image-set were performed by [Mark Quinlan] and [Aaron Ceross], with all further analysis performed by [Mark Quinlan]. The first draft of the manuscript was written by [Mark Quinlan] and all authors commented on previous versions of the manuscript. All authors read and approved the final manuscript.

The authors declare that no funds, grants, or other support were received during the preparation of this manuscript, and have no relevant financial or non-financial interests to disclose.

\bibliography{sn-bibliography}

\end{document}